\documentclass{PoS}

\title{Hadron spectrum of QCD with one quark flavor}

\ShortTitle{Hadron spectrum of QCD with one quark flavor}

\author{Federico Farchioni{}\speaker{}, Gernot M\"unster, %
          Tobias Sudmann$^\thefootnote$, Ja\"ir Wuilloud\\
        Universit\"at M\"unster, Institut f\"ur Theoretische Physik,\\
        Wilhelm-Klemm-Strasse~9, D-48149~M\"unster, Germany\\
        E-mail: \email{farchion@mail.desy.de, tobias.sudmann@uni-muenster.de}}

\author{Istv\'an Montvay\\
        Deutsches Elektronen-Synchrotron DESY, %
        Notkestr.~85, D-22603~Hamburg, Germany}

\author{Enno\,E. Scholz\\
        Physics Department, Brookhaven National Laboratory, %
        Upton,~NY~11973, USA}

\abstract{
The hadron spectrum of one flavor QCD is studied by Monte Carlo
simulations. The Symanzik tree-level-improved Wilson action is used
for the gauge field and the Wilson action for the
fermion. The theory is simulated by a polynomial hybrid Monte Carlo
algorithm (PHMC).  The mass spectrum of hadronic bound states is
investigated at two different lattice spacings: $a \simeq 0.37\, r_0$
and $a \simeq 0.27\, r_0$, corresponding to $\simeq 0.19 {\rm\, fm}$
and $\simeq 0.13 {\rm\, fm}$ in ordinary QCD. The lattice extension is
fixed to $L \simeq 4.4\, r_0\; (\simeq 2.2 {\rm\, fm})$. The lightest
simulated quark mass corresponds to a pion with mass $\sim 270$~MeV.
Properties of the theory are analyzed by making use of
the ideas of partially quenched chiral perturbation theory (PQChPT).
The symmetry of the single flavor theory can be
artificially enhanced by adding extra valence quarks, which can be
interpreted as $u$ and $d$ quarks. Operators in the valence pion
sector can be built.  Masses and decay constants are analyzed by using
PQChPT formulae at next-to-leading order.

}

\FullConference{The XXV International Symposium on Lattice Field Theory\\
		 July 30-4 August 2007\\
		 Regensburg, Germany}
\begin{document}

\newcommand{\be}{\begin{equation}}
\newcommand{\ee}{\end{equation}}
\newcommand{\bea}{\begin{eqnarray}}
\newcommand{\eea}{\end{eqnarray}}
\newcommand{\half}{\frac{1}{2}}
\newcommand{\rar}{\rightarrow}
\newcommand{\lar}{\leftarrow}
\newcommand{\LCB}{\raisebox{-0.3ex}{\mbox{\LARGE$\left\{\right.$}}}
\newcommand{\RCB}{\raisebox{-0.3ex}{\mbox{\LARGE$\left.\right\}$}}}
\newcommand{\LSB}{\raisebox{-0.3ex}{\mbox{\LARGE$\left[\right.$}}}
\newcommand{\RSB}{\raisebox{-0.3ex}{\mbox{\LARGE$\left.\right]$}}}
\newcommand{\tr}{{\rm Tr}}
\newcommand{\I}{\ensuremath{\mathrm{i}\,}}
\newcommand{\E}{\ensuremath{\mathrm{e}\,}}
\section{Introduction and motivation}

QCD with one flavor of quarks ($N_f=1$ QCD) radically differs from QCD with 
two or more flavors due to the absence 
of a chiral symmetry: the abelian symmetry of the one flavor theory is washed out 
at the quantum level by the Adler-Bell-Jackiw anomaly. Only
a vector symmetry survives, related to the conservation of the quark number.
As a consequence of this, the main features of the phase structure and mass spectrum 
of the single flavor theory strongly deviate from the familiar picture, affected by the spontaneous 
breaking of the non-abelian chiral symmetry, of ordinary QCD.
These unphysical features of $N_f=1$ QCD explain the little attention
reserved to this kind of setup in past simulations
(however with the exceptions of~\cite{AlexandrouThermodyn} and~\cite{DeGrandHoffmannSchaeferLiu}).

This situation has changed in recent years, mainly due to the works of M.~Creutz
drawing the attention of the lattice community to open problems in the physical ($N_f>1$) 
theory~\cite{Creutz:UpMass,Creutz:CP}.
Since these aspects are not directly related to the spontaneous breaking of the chiral symmetry,
they find an equivalent in the single flavor theory. The latter represents therefore a simple setup 
for their investigation.

One question raised by Creutz~\cite{Creutz:UpMass}, having a relevant phenomenological impact, 
is whether it is possible to define in an unambiguous way the case where {\em one} quark
(say the $u$ quark) becomes massless. The arguments
against an unique definition of the massless limit~\cite{Creutz:UpMass}
essentially rest upon the $U(1)$ anomaly and should therefore hold {\em a fortiori} 
for the one flavor theory.
A second aspect is the possibility of a spontaneous breaking of CP in QCD for special choices 
of the quark masses, conjectured for the first time
by Dashen~\cite{Dashen}.
According to the Vafa-Witten theorem~\cite{VafaWitten} a prerequisite 
for the spontaneous breaking of a discrete symmetry is a non positive fermion measure,
which in $N_f=1$ QCD is possible for negative quark masses. The  
transition line is indeed expected to be located~\cite{Creutz:CPNf1} 
on the negative real quark mass axis in the extended complex parameter space.
In the case of the multi-flavor theory, the transition is excluded for physical
values of the quark mass, but its nearby presence can nevertheless
affect numerical simulations on the lattice~\cite{Creutz:CP}. 
So the main features of this transition are not of academic interest only.

Another intriguing aspect of one flavor QCD, emerging from string theory, is the connection 
with the ${\cal N}$=1 supersymmetric Yang-Mills theory (SYM).
The equivalence of the two theories in the bosonic sector~\cite{ArmShiVen} can be proven at the planar level of a
particular large $N_c$ limit (orientifold large $N_c$ limit) preserving balance between
fermionic and bosonic degrees of freedom.
Relics of SUSY are therefore expected in $N_f=1$ QCD (with $N_c=3$).
A prediction of the orientifold equivalence~\cite{ArmShiVen_Cond}, already studied in the 
literature~\cite{DeGrandHoffmannSchaeferLiu}, 
concerns in particular the size of the quark condensate. 

Another important place where relics of SUSY in $N_f=1$ QCD can be investigated, 
considered in more detail in this contribution, is the low-lying
bound-state spectrum~\cite{OurNf1}. In SYM the mass patterns are strongly constrained by SUSY.
In particular low-energy models~\cite{VenezianoYankielovicz} predict a low-lying 
chiral supermultiplet including two scalar particles with opposite 
parity.\footnote{For recent lattice simulations of ${\cal N}$=1 SYM, see~\cite{Ward_SYM,FaPe}.}
In $N_f=1$ QCD these two particles can be easily identified with the $\eta$ 
and the $\sigma$ meson (the former picking up a mass through the anomaly). 
On the basis of the planar equivalence, their mass ratio including $O(1/N_c)$ corrections is expected 
to be $m_\sigma/m_\eta =N_c/(N_c-2)$~\cite{ArmoniImeroni}.

The study of the mass spectrum of hadronic states 
requires reasonably large physical volumes, in order to be able to accommodate 
the bound-states in the finite box,
and small (positive) quark masses.  High statistics is required
for a precise determination of the disconnected quark diagrams needed for the 
scalar meson masses, which are characterized 
by a high level of noise. High statistics is also important
for the computation of the glueball masses.
We apply here the Wilson lattice fermion action which 
has recently been
shown~\cite{qq+q,CERN,QCDSF,ETMC} to be well
suited for such an investigation.
Preliminary results with Stout-smeared links~\cite{Stout} in the Wilson
fermion action will be also presented.
Following~\cite{ETMC} we apply in the gauge sector 
the tree level improved Symanzik action~(tlSym).

The present exploratory study has been performed  on $12^3 \cdot 24$ and
$16^3 \cdot 32$ lattices with a lattice spacing corresponding in QCD units to $a \simeq 0.19\,{\rm fm}$
and $a \simeq 0.13\,{\rm fm}$, respectively. 
(We use the Sommer parameter~\cite{Sommer} $r_0$
for setting the scale, fixed at the conventional value 
$r_0 \equiv 0.5\,{\rm fm}$.)
For the future we plan to run simulations
closer to the continuum limit.

As already mentioned, the sign of the quark determinant is an important issue in $N_f=1$ QCD
(in particular, a negative determinant triggers the CP-violating 
phase transition). In the continuum, the fermion determinant is positive for positive quark mass.
With Wilson lattice fermions for small quark masses, it can become negative 
due to quantum fluctuations.
In most of our simulations the quark mass is large enough to prevent sign changes
and the occurrence of a negative determinant is a rare event. For the lightest simulated
quark masses however 
the sign of the quark determinant may potentially play a role and its impact
in the hadron spectrum  must be checked.
In our simulations we could reach quite small quark masses down to $m_q \simeq 12{\rm\,MeV}$ 
($m_q r_0 \simeq 0.03$), corresponding to a pion mass
$m_\pi \simeq 270{\rm\,MeV}$.

As we have argued in~\cite{OurNf1},
it is useful to embed the $N_f=1$ QCD theory in a {\em partially quenched}
theory with additional quark flavors. A particularly symmetric choice
consists in taking the  ($N_{V}$) {\em valence} quark flavors degenerate with
the {\em sea} quark: in this case the combined sea and valence sector is characterized
by an exact ${\rm SU}(N_{V}+1)$ flavor symmetry. 
In this fictitious multi-flavor theory a PCAC quark mass can be 
naturally defined. We take this quantity as an operative definition of the quark mass for the
(unitary) one-flavor theory. Also, a partially quenched
chiral perturbation theory (PQChPT) can be set-up, exactly as in the $N_f>1$ case.
The latter reduces to an effective theory of the $\eta$ meson in the 
unitary sector without valence quarks.
The predictions of this PQChPT will be compared against our
numerical data.

The plan of this contribution is as follows: in the next section the
partially quenched viewpoint is introduced and PQChPT is considered
for it.  In Section~\ref{sec:sim} some information on the 
simulation algorithm and on the computation of the sign of the determinant 
are given.  Section \ref{sec:spec} is devoted to
the presentation of our numerical results on the hadron spectrum, while
Section~\ref{sec:pq} discusses the partially quenched data.  The
last section contains summary and outlook.

\section {Partially quenched QCD}

The symmetry of the one flavor theory can be artificially enhanced by
adding extra valence quarks which are {\em quenched}, namely not taken
into account in the Boltzmann-weight of the gauge configurations by
their fermion determinants.  
A theoretical description of the resulting partially quenched theory can be obtained
through the introduction of ghost quarks~\cite{Morel}. In this method
the functional integral over the ghost quark fields
$\tilde\psi$ cancels the fermion determinant of the valence quarks $\psi_V$,
\begin{eqnarray}
 && \int\mathcal{D}A\,\mathcal{D}[\psi_S\bar\psi_S]\,\,\mathcal{D}[\psi_V\bar\psi_V]\,
        \mathcal{D}[\tilde\psi\bar{\tilde\psi}]\;
     \mathrm{e}^{- S_\mathrm{g} - \bar\psi_S(\gamma_\mu D_\mu+m_S)\psi_S
                   - \bar\psi_V(\gamma_\mu D_\mu+m_V)\psi_V
                   - \bar{\tilde\psi}(\gamma_\mu D_\mu+m_V)\tilde\psi}\nonumber \\
&=& \int\mathcal{D}A\; \mathrm{e}^{-S_\mathrm{g}}\;
          \frac{\det(\gamma_\mu D_\mu+m_V)}{\det(\gamma_\mu D_\mu+m_V)}
          \det(\gamma_\mu D_\mu+m_S)\,,
\end{eqnarray}
and only the determinant of the sea ($S$) quark remains in the measure.  In principle,
one might consider any number of quenched valence quarks with any mass
values. In our approach we take two valence quarks $u$ and $d$ with masses $m_V$ and one sea
quark $s$ with mass $m_S$. For our purpose the case of degenerate
valence and sea quark mass $m_V=m_S$ is particularly convenient (which
is admittedly an unconventional kind of partially quenching). 
Observe that in this symmetric setup the exact number of valence quarks $N_V$ is immaterial,
so our position $N_F\equiv N_V+N_f=2+1$ is just suggested by 
analogy with the case realized in nature. (Of course, in order to be able to build
bound states containing two different quark flavors as mesons and nucleons, one needs 
$N_V\geq 1$.)


At the point of vanishing quark masses (see~below) the generic partially
quenched theory has a graded ${\rm SU}(N_F|N_V)_L\,\otimes\, {\rm SU}(N_F|N_V)_R$ 
symmetry, which is broken spontaneously into a
``flavor'' symmetry ${\rm SU}(N_F|N_V)$, also valid for non-vanishing 
degenerate quark masses. The ${\rm SU}(N_F)$ subgroup represents the flavor symmetry in the
combined sea and valence quark sectors. The latter symmetry implies that the hadronic bound states
appear in exactly degenerate SU($N_F$) multiplets for  $m_V=m_S$.

In particular, this extended theory contains a degenerate octet of pseudoscalar mesons 
(``pions'' $\pi^a$,  $a=1,\ldots,8$)
satisfying an ${\rm SU}(3)$-symmetric PCAC relation. Considering the
divergence of the axial-vector current $A^a_{x\mu}$ and pseudoscalar
density $P^a_x$ we can define the bare {\em PCAC
  quark mass} in lattice units as usual
\begin{equation}
am_{\rm\scriptscriptstyle PCAC} \equiv
\frac{\langle \partial^\ast_\mu A^+_{x\mu}\, P^-_y \rangle}
{2\langle P^+_x\, P^-_y \rangle} \, .
\end{equation}
Here the indices $+$ and $-$ refer to the ``charged'' components
corresponding to $\lambda_a \pm {\rm i}\lambda_b$ (with
$\lambda_{a,b}$ some off-diagonal Gell-Mann matrices) and
$\partial^\ast_\mu$ denotes the backward lattice derivative.  Due to
the exact SU(3)-symmetry, the renormalized quark mass corresponding to
$m_{\rm\scriptscriptstyle PCAC}$ can be defined by an SU(3)-symmetric
multiplicative renormalization:
\begin{equation}\label{eq:pcac}
m^{\rm\scriptscriptstyle R}_{\rm\scriptscriptstyle PCAC} =
\frac{Z_A}{Z_P} m_{\rm\scriptscriptstyle PCAC} \ .
\end{equation}

As we will confirm numerically in sec.~\ref{sec:pq},   the masses of
the ``pions'' can be made to vanish by suitably tuning the bare quark mass on the lattice.
In this situation the renormalized quark mass~(\ref{eq:pcac}) vanishes, too. We stress here 
that the pions are not particles in the physical 
spectrum of the theory. Nevertheless their properties 
as mass and decay constant are well
defined quantities which can be computed on the lattice. 
The same applies for the PCAC quark mass $m^{\rm\scriptscriptstyle R}_{\rm\scriptscriptstyle PCAC}$ 
which can be therefore 
regarded as a potential candidate for a definition of the quark mass in
this theory.

\subsection{Chiral perturbation theory}

The dependence of pion properties upon the quark
masses can be determined in partially quenched chiral perturbation theory
(PQChPT) \cite{BG,SharpePQ}. The  effect of the finite lattice spacing
$a$ can be also included~\cite{ShaSi,LeeSha,Rupak-Shoresh,Aoki,BRS}.  
The pseudo-Goldstone fields are parameterized by a graded matrix
\begin{equation}
U(x) = \exp \left(\frac{\mathrm{i}}{F_0} \Phi(x) \right)\,,
\end{equation}
which in our case is in the supergroup ${\rm SU}(3|2)$. (The normalization
of $F_0$ is chosen such that its phenomenological value is $\simeq
86{\rm\,MeV}$.) The commuting elements of the graded matrix $\Phi$
represent the pseudo-Goldstone bosons made from a quark and an
anti-quark with equal statistics, while the anticommuting elements 
represent pseudo-Goldstone fermions which are built from one
fermionic quark and one bosonic quark.  The supertrace of $\Phi$ has
to vanish, which can be implemented by a suitable choice of generators
\cite{Sharpe-Shoresh}.

We have calculated both the masses and decay constants of the
pseudo-Goldstone bosons in next-to-leading order of partially quenched
chiral perturbation theory along the lines of
Ref.~\cite{Sharpe-Shoresh}, including ${\cal O}(a)$ lattice artifacts
\cite{Rupak-Shoresh}.  The quark masses enter the expressions in the
combinations
\begin{equation}
\chi_V = 2 B_0\, m_V\,,\quad \chi_S = 2 B_0\, m_S\,,\quad
\chi_{\rm\scriptscriptstyle PCAC}
= 2 B_0\, m^{\rm\scriptscriptstyle R}_{\rm\scriptscriptstyle PCAC}
\end{equation}
with the usual leading order low-energy constant $B_0$; the lattice spacing
enters in the combination
\begin{equation}
\rho = 2 W_0\, a,
\end{equation}
where $W_0$ is another, lattice-specific, low-energy constant. We have
calculated the masses of the pions and mixed mesons (degenerate in the special case $m_V=m_S$). 
The next-to-leading-order expression in terms of the (renormalized) PCAC
quark mass is
\begin{eqnarray}\label{eqchpt}
m_{\pi}^2 = \chi_{\rm\scriptscriptstyle PCAC}
&+& \frac{\chi_{\rm\scriptscriptstyle PCAC}^2}{16 \pi^2 F_0^2}
\ln \frac{\chi_{\rm\scriptscriptstyle PCAC}}{\Lambda^2}
+ \frac{8}{F_0^2}
\Big[ (2L_8-L_5+2L_6-L_4) \chi_{\rm\scriptscriptstyle PCAC}^2 \nonumber\\
&&\qquad +\, (W_8+W_6-W_5-W_4-2L_8+L_5-2L_6+L_4)
\chi_{\rm\scriptscriptstyle PCAC}\rho\Big]\,,
\end{eqnarray}
where the usual next-to-leading order low-energy parameters $L_i$ appear, together with
additional ones ($W_i$) describing lattice artifacts.
For the decay constant we obtain in this case
\begin{equation}\label{eqchpt2}
F_\pi = F_0 \cdot \Big\{
1 - \frac{\chi_{\rm\scriptscriptstyle PCAC}}{32 \pi^2 F_0^2}
\ln \frac{\chi_{\rm\scriptscriptstyle PCAC}}{\Lambda^2}
+ \frac{8}{F_0^2}
\big[ (L_5+L_4) \chi_{\rm\scriptscriptstyle PCAC}
+ (W_5+W_4-L_5-L_4) \rho \big] \Big\}\,.
\end{equation}
Observe that as expected the results are independent of $N_V$. 
In particular, calculating the quantities in this model with
$N_V=1$, which corresponds to a representation the supergroup ${\rm SU}(2|1)$, 
reproduces (\ref{eqchpt}) and (\ref{eqchpt2}).

The analysis can be extended by relating the pion mass to the mass of
the ``physical'' $\eta$. The inclusion of the singlet can be
achieved by relaxing the constraint of a vanishing supertrace
\cite{BG,Sharpe-Shoresh}, and associating it with the field
\begin{equation}
\Phi_0 (x) = {\rm sTr}\, \Phi(x).
\end{equation}
The effective Lagrangian then contains additional terms depending on
$\Phi_0$:
\begin{equation}
\Delta \mathcal{L} =
\alpha \partial_{\mu} \Phi_0 \partial_{\mu} \Phi_0 + m_{\Phi}^2 \Phi_0^2
+ \mathcal{O}(\Phi_0^3)\,,
\end{equation}
where $\alpha$ and $m_{\Phi}$ are free parameters in this context.  We
will use in the following the leading order expression
for the mass of the $\eta$, which reads
\begin{equation}\label{eq:eta}
m_{\eta}^2 =
\frac{m_{\Phi}^2 + \chi_{\rm\scriptscriptstyle PCAC}}{1 + \alpha}.
\end{equation}
Our numerical results for $m_{\eta}$ allow to determine $\alpha$ and
$m_{\Phi}$ (see Section~\ref{sec:pq}).

\section{Simulation}
\label{sec:sim}

For the SU(3) gauge sector we apply 
the tree-level improved Symanzik (tlSym) action~\cite{WeiszWohlert}
including planar rectangular $(1\times 2)$ Wilson loops:
\be\label{eq01}
S_g = \beta\sum_{x}\left(c_{0}\sum_{\mu<\nu;\,\mu,\nu=1}^4
\left\{1-\frac{1}{3}\,{\rm Re\,} U_{x\mu\nu}^{1\times 1}\right\}
+c_{1}\sum_{\mu\ne\nu;\,\mu,\nu=1}^4
\left\{1-\frac{1}{3}\,{\rm Re\,} U_{x\mu\nu}^{1\times 2}\right\}
\right) \ ,
\ee
with $c_1=-1/12$  and normalization condition $c_{0}=1-8c_{1}$.
The fermionic part of the lattice action is the simple (unimproved) Wilson action.
With the goal of improving the stability of the Monte Carlo evolution at small quark masses, we
also started simulations with Stout-smeared links~\cite{Stout} in the hopping matrix (see below).


The update algorithm is a Polynomial Hybrid Monte Carlo algorithm (PHMC)~\cite{ForcrandTakaishi,FrezzottiJansen} 
allowing the simulation of an odd number of fermion species.
The present version~\cite{MontvayScholz} is based on a two-step polynomial
approximation of the inverse fermion matrix with stochastic
correction in the update chain: a {\em sequence} of PHMC trajectories is followed
by a Metropolis accept-reject step with a higher precision polynomial.
The polynomial approximation scheme and the stochastic correction
in the update chain are taken over from the two-step multi-boson algorithm of Ref.~\cite{Montvay:tsmb}.
A correction factor $C[U]$ in the measurement is associated with configurations
for which eigenvalues of the (squared Hermitian) fermion matrix $Q^2[U]$ 
lie outside the validity interval of the polynomial approximation. 
We refer to~\cite{OurNf1} for more details 
on the algorithmic setup.

As mentioned in the Introduction, the sign $\sigma[U]$ of the
fermion determinant $\det Q[U]$ has also to be included in the
reweighting of the configurations. The expectation
value of a quantity $A$ is therefore given by
\be\label{eq08}
\langle A \rangle = \frac{\int [dU]\:\sigma[U]\,C[U]\,A[U]}
                         {\int [dU]\:\sigma[U]\,C[U]}  \ .
\ee

For the computation of the sign $\sigma[U]$ we applied two methods.
In the first we studied the {\em spectral flow} of the Hermitian fermion matrix~\cite{EdwardsFlow}. 
For the $\kappa$-dependent computation of the low-lying eigenvalues of the Hermitian fermion matrix 
$Q[U]$ we followed in this case Ref.~\cite{KalkreuterSimma}. Alternatively, we computed
the (complex) spectrum of the non-Hermitian matrix concentrating on the lowest real eigenvalues:
sign changes are signaled by negative real eigenvalues. We applied
the ARPACK Arnoldi routines~\cite{Arpack} on a transformed Dirac operator.
The (polynomial) transformation was tuned such that the real eigenvalues were 
projected outside the ellipsoidal bulk containing the whole eigenvalue spectrum~\cite{Saad}.
This allows for an efficient computation of the real eigenvalues~\cite{Neff}.
This latter method, on which we will rely in the future,  
delivers unambiguous results and can be simply automatized.



\begin{table}
\begin{center}
  \caption{\label{tabruns}{ Summary of the runs: $12^3\cdot 24$ and
      $16^3\cdot 32$ lattices have lowercase and uppercase labels,
      respectively.  The bar indicates runs with Stout-link in the fermion action 
      (see text).\vspace{\baselineskip}}}

%
\renewcommand{\arraystretch}{1.2}
\begin{tabular}{c*{6}{|c}}
 \multicolumn{1}{c|}{}&
 \multicolumn{1}{c|}{$\beta$} &
 \multicolumn{1}{c|}{$\kappa$} &
 \multicolumn{1}{c|}{$N_{\textrm{conf}}$} &
 \multicolumn{1}{c|}{${\textrm{plaquette}}$} &
 \multicolumn{1}{c|}{$\tau_{\textrm{plaq}}$} &
 \multicolumn{1}{c}{$r_0/a$}
 \\
\hline\hline
 $a$ & 3.80 & 0.1700 & 5424 & 0.546041(66) & 12.5 & 2.66(4)
 \\ \hline
 $b$ & 3.80 & 0.1705 & 3403 & 0.546881(46) & 4.6  & 2.67(5)
 \\ \hline
 $c$ & 3.80 & 0.1710 & 2884 & 0.547840(67) & 7.6  & 2.69(5)
 \\ \hline\hline
 $A$ & 4.00 & 0.1600 & 1201 & 0.581427(36) & 4.3  & 3.56(5)
 \\ \hline
 $B$ & 4.00 & 0.1610 & 1035 & 0.582273(36) & 4.1  & 3.61(5)
 \\ \hline
 $C$ & 4.00 & 0.1615 & 1005 & 0.582781(32) & 3.3  & 3.73(5)
 \\ \hline\hline
 $\bar A$ & 4.00 & 0.1440 & 5600 & 0.577978(23) & 9.7  & 3.74(3)
 \\ \hline
 $\bar B$ & 4.00 & 0.1443 & 5700 & 0.578167(28) & 11.3  & 3.83(5) 
 \\ \hline\hline
\end{tabular}
\end{center}
\end{table}

\subsection{Simulation details}

We performed simulations on a $12^3\cdot 24$ lattice with $\beta=3.8$ and 
on a $16^3\cdot 32$ with $\beta=4.0$. 
Information regarding the generated sets of configurations are reported in 
Table~\ref{tabruns}. 

The sequences consisted of 3--6 PHMC individual trajectories.
The precision of the first step of polynomial approximations was
tuned such that the acceptance of the PHMC trajectories was about 0.80--0.85.
The same acceptance was required for the Metropolis test 
by tuning the total length of the trajectory (1.5--1.8).
This resulted in a relatively high total acceptance of 0.64--0.72.
Optimization of the parameters of PHMC turned out to have a substantial 
impact on the integrated autocorrelation times of the average plaquette.

In the case of a Stout-link we consider one step of isotropic 
smearing with $\rho_{\mu\nu}=\rho=0.15$, $\mu,\nu=1,\ldots, 4$. The Stout-smearing has 
in general the beneficial effect, compared to the unsmeared action, of reducing the fluctuations 
of the smallest eigenvalue of the (squared) hermitian matrix, with the result that 
less exceptional configurations are observed. This allowed us to obtain smooth simulations 
down to quite small pion masses $m_\pi\simeq 270$~MeV.

Taking the values of $r_0/a$ at the highest $\kappa$'s
for the runs at $\beta=3.8$ and $\beta=4.0$ and fixing $r_0=0.5{\rm\: fm}$ by definition
we obtain $a = 0.186{\rm\, fm}$ and $a = 0.134{\rm\: fm}$, respectively.
The extensions of the $12^3$ and $16^3$ lattices 
are roughly constant: $L = 2.23{\rm\: fm}$ and $L = 2.14{\rm\: fm}$.
(The Stout-smearing leaves $r_0/a$ essentially unchanged.)


For runs $b$, $c$, $\bar{A}$ and $\bar{B}$ there are cases where 
the eigenvalues of the fermion matrix are outside the
approximation interval $[\epsilon,\lambda]$ and therefore $C[U]\neq 1$.
In run $c$ in particular there are 167 of such configuration out of 2884, 26 
of them with negative sign. However, even in this case the average value
$\sigma[U] C[U]$ is very near to one: 0.9842.
The effect of the correction factors turns out to be quite weak in the case of
the average plaquette and of $r_0/a$: the effect on the
average value of $r_0/a$ is only in the fifth digit (whereas the
statistical error is in the third digit). This is not the case
for low energy quantities as the low-lying hadron masses (see in the following).


\section{Hadron spectrum}
\label{sec:spec}

\subsection{Mesons}

For the meson states we consider the simplest
interpolating operators in the pseudoscalar and scalar sectors:
\bea
\eta   (0^-):   \quad P(x)&=& \bar\psi(x)\gamma_5\psi(x)~, \\
\sigma (0^+): \quad S(x)&=& \bar\psi(x)\psi(x)~.
\eea
Corresponding states in the QCD spectrum 
are the $\eta^\prime(958)$ and $f_0(600)$ 
(or $\sigma$).
In the case of the pseudoscalar mesons, invariance under the flavor
group plays a special role when comparing with QCD states  because of
the  U(1) axial anomaly. 

The disconnected diagrams of the hadron correlators of $\eta$ and $\sigma$ 
were computed by applying stochastic sources with complex $Z_2$ noise and spin
dilution. The method was already applied to the case of lattice SYM~\cite{FaPe}.
In order to optimize the computational load, also considering
autocorrelations, we analyzed typically every fifth configuration, 
with 20 stochastic estimates each. The resulting statistics 
is $400-600$ on the smaller lattice and $\sim 200$ on the larger one.

\begin{table}[t]
\begin{center}
\caption{\label{tab:hadmass}
 Results for light hadron masses in $N_f=1$ QCD
 (the result with the asterisk has been obtained with higher
  statistics: 4900). Note that the glueball masses were obtained at small time separations
and hence could be overestimated (see also text).
}\vspace{\baselineskip}
%
\renewcommand{\arraystretch}{1.2}
\begin{tabular}{c*{4}{|c}}
 \multicolumn{1}{c|}{}&
 \multicolumn{1}{c|}{$am_{\eta}$} &
 \multicolumn{1}{c|}{$am_{\sigma}$} &
 \multicolumn{1}{c|}{$am_{0^{++}}$} &
 \multicolumn{1}{c}{$am_{\Delta}$} 
 \\
\hline\hline
 $a$ & 0.462(13)     & 0.660(39)       & 0.777(11)     & 1.215(20)
 \\ \hline
 $b$ & 0.403(11)     & 0.629(29)       & 0.685(10)     & 1.116(38)
 \\ \hline
 $c$ & 0.398(28)     & 0.584(55)       & 0.842(16)     & 1.204(57)
 \\ \hline\hline
 $A$ & 0.455(17)     & 0.607(57)       & 1.083(79)     & 1.006(15)
 \\ \hline
 $B$ & 0.380(18)     & 0.554(52)       & 1.032(66)     & 0.960(15)
 \\ \hline
 $C$ & 0.316(22)     & 0.613(67)       & 0.943(41)$^*$ & 0.876(26)
 \\ \hline\hline
\end{tabular}
\end{center}
\end{table}

\subsection{Baryons}

The simplest interpolating field in the baryon sector containing just one quark field is
\be\label{eq:rarita-schw}
 {\Delta_i}(x)\:=\:\epsilon_{abc}[\psi_a(x)^TC\gamma_i\psi_b(x)]\psi_c(x)\ .
\ee
The low lying hadron state interpolated by the above operator is 
expected to have spin $3/2$ and positive parity $(\frac{3}{2}^+)$.
This corresponds to the $\Delta^{++}(1232)$ of QCD if our dynamical
fermion is interpreted as an $u$ quark
(the $\Omega^-$ baryon is more appropriate for larger quark masses).

A difficulty arises since the Rarita-Schwinger spinor~(\ref{eq:rarita-schw}) also contains a spin $1/2$ component.
We extract the wanted spin 3/2 component by projection~\cite{regina}:
\be
G_{3/2}(t)=\frac{1}{6}{\rm Tr}
 \left[ G_{ji}(t)\gamma_j\gamma_i+G_{ii}(t)\right]
\ , \quad 
G_{ji}(t)=\sum_{\vec{x}}
\left\langle \Delta_j(\vec{x},t)\bar\Delta_i(0)\right\rangle \ .
\ee
Since the baryon correlator does not contain disconnected diagrams,
our full statistics could be taken for the computation of the masses in this case,
namely 3000--4000 on the smaller lattice and $\sim 1000$ on the larger one.

\begin{figure}[t]
\begin{center}
  \includegraphics[angle=-90,width=.5\linewidth]{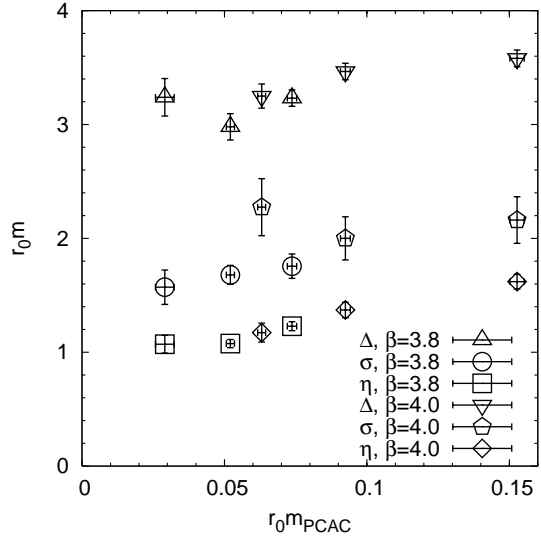}
  \caption{\label{fig:hadmass}
    The mass of the lightest physical particles in one-flavor QCD as
    a function of the bare PCAC quark mass.  The masses are multiplied by
    the scale parameter $r_0$ in order to obtain dimensionless
    quantities.  Open and full symbols refer to $\beta=3.8$ and
    $\beta=4.0$, respectively.}
\end{center}
\end{figure}

\subsection{Glueballs}

Spin~0 states are also projected by purely gluonic operators.  These
are the glueballs, a well known object of investigation in lattice
QCD.  In particular the $0^{++}$ glueball has the same quantum numbers
as the $\sigma$ meson.  In this first investigation we neglect
possible mixings between the two states and consider only diagonal
correlators.  

We used the single spatial plaquette to obtain the mass
of the $0^{++}$ ground state.  To increase the overlap of the operator
with this state we performed APE smearing~\cite{APE} and also applied variational
methods~\cite{VarSm} to obtain optimal glueball operators from linear combinations
of the basic operators.


\subsection{Results}

The results for the hadron masses (only available for the runs without Stout-smearing) are
reported in lattice units in  Table~\ref{tab:hadmass}. 
In Fig.~\ref{fig:hadmass} the hadron masses are plotted 
as a function of the bare PCAC quark mass
$m_{\rm \scriptscriptstyle PCAC}$~(\ref{eq:pcac}) defined in the partially quenched picture.
Since we use physical units here, results from the 
two lattice spacings can be compared. The scaling is satisfactory  
for the case of $\eta$, whose mass could be computed with the best accuracy.
The determination of the $\sigma$ meson mass seems to require large statistics.

The effect of the sign of the determinant in the hadron spectrum
was investigated by computing the masses with or without the inclusion
of the sign factor in the reweighting procedure.
Only in the case of our run at the lightest quark mass, run~$c$,  
a sizeable effect can be observed: here the sign of the determinant pushes up
the masses by $7-10$\%.

We observe that our statistics is not large enough to obtain an accurate
estimate of the glueball masses. In particular, the results reported in 
Table~\ref{tab:hadmass} could be overestimated. Indeed, due to the high level of noise,
large time-separations could not be included in the determinations; it is therefore
possible that the latter are contaminated by excited states.
In order to enhance the statistics 
we decided to store the gauge configuration more frequently 
(as was already applied for the continuation of run $C$).

\section{Partially quenched analysis}
\label{sec:pq}

\begin{table}[t]
\begin{center}
  \caption{\label{tab:valhad}
    The PCAC quark mass $m_{\rm\scriptscriptstyle PCAC}$, the pion
    mass $m_\pi$ and non-renormalized decay constant $f_\pi$, and the nucleon mass
    $m_N$ in lattice units (only for the runs without Stout-smearing).}\vspace{\baselineskip}
%
\renewcommand{\arraystretch}{1.2}
\begin{tabular}{c*{4}{|l}}
 \multicolumn{1}{c|}{}&
 \multicolumn{1}{c|}{$am_{\rm\scriptscriptstyle PCAC}$} &
 \multicolumn{1}{c|}{$am_\pi$} &
 \multicolumn{1}{c|}{$af_\pi$} &
 \multicolumn{1}{c}{$am_N$} 
 \\
\hline\hline
 $a$ &   0.02771(45) &   0.3908(24) &  0.1838(11) &  1.0439(54)
 \\ \hline
 $b$ &   0.01951(39) &   0.3292(25) &  0.1730(15) &  0.956(27)
 \\ \hline
 $c$ &   0.0108(12)  &   0.253(10)  &  0.156(10)  &  1.011(51)
 \\ \hline\hline
 $A$ &   0.04290(36) &   0.4132(21) &  0.1449(9)  &  0.9018(44)
 \\ \hline    
 $B$ &   0.02561(31) &   0.3199(22) &  0.1289(10) &  0.7978(53)
 \\ \hline
 $C$ &   0.01700(30) &   0.2635(24) &  0.1188(12) &  0.734(10)
 \\ \hline\hline
 $\bar A$ &   0.01532(34) & 0.2316(49) &   0.09747(15)     & 
 \\ \hline
 $\bar B$ &   0.00886(75) & 0.1994(74) &   0.0852(49)      & 
 \\ \hline\hline
\end{tabular}
\end{center}
\end{table}

\begin{figure}[t]
\begin{center}
  \includegraphics[angle=-90,width=.5\linewidth]{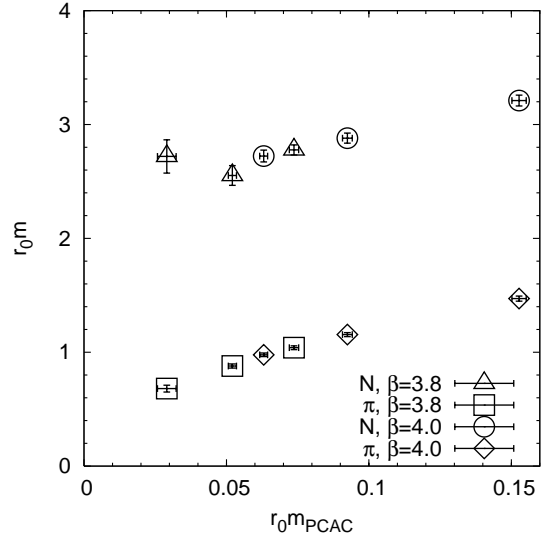}
  \caption{\label{fig:valhad}
    The mass of the valence pion and nucleon as a function of the bare
    PCAC quark mass.  Open and full symbols refer to $\beta=3.8$ and
    $\beta=4.0$, respectively.}
\end{center}
\end{figure}

The results for the partially quenched sector are collected in
Table~\ref{tab:valhad} and shown in Figure~\ref{fig:valhad}.
This also includes the nucleon mass (only for the runs without Stout-smearing).

\begin{figure}[t]
  \begin{center}
  \includegraphics[angle=-90,width=.45\linewidth]{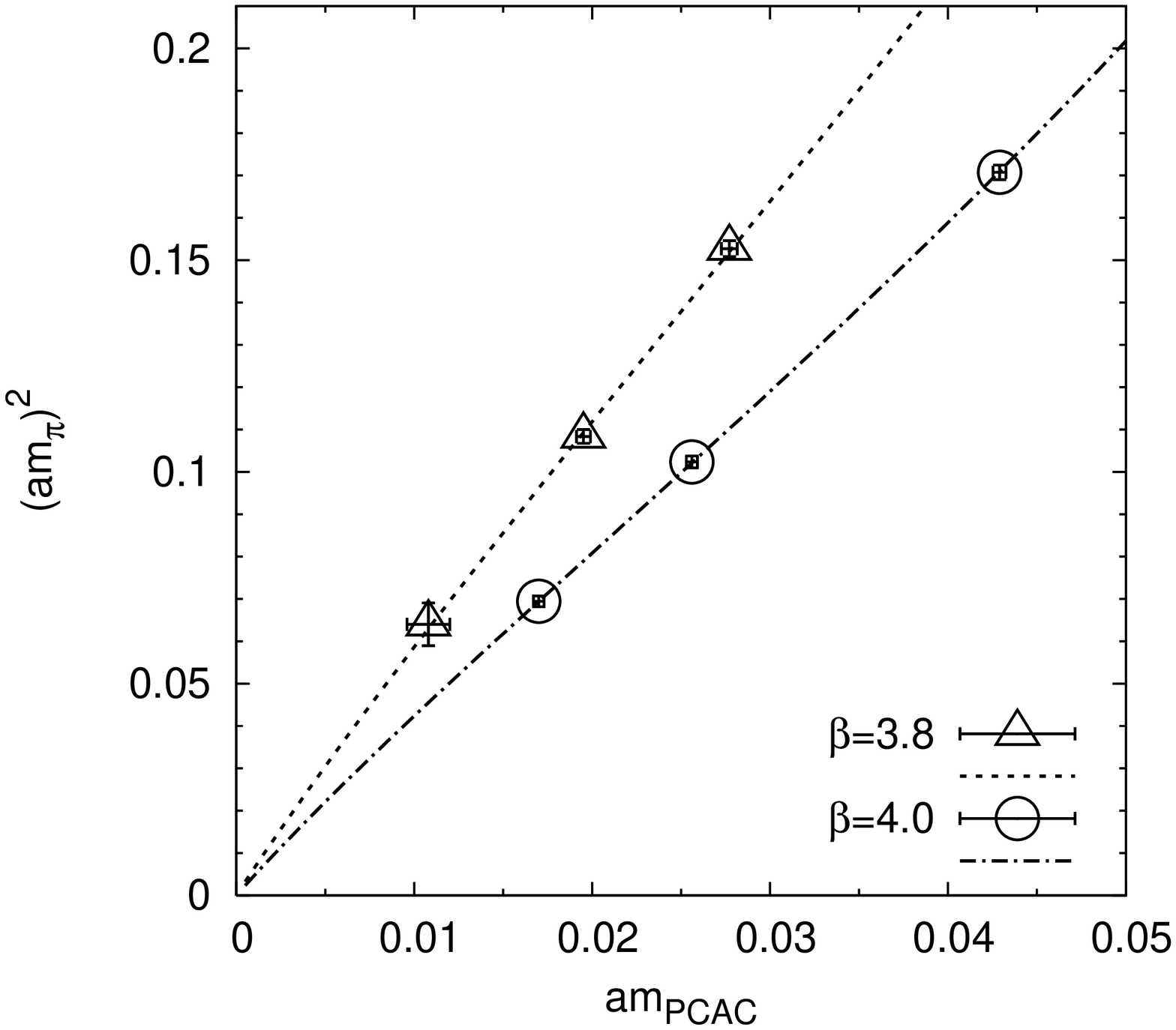}
  \hspace{.05\linewidth}
  \includegraphics[angle=-90,width=.45\linewidth]{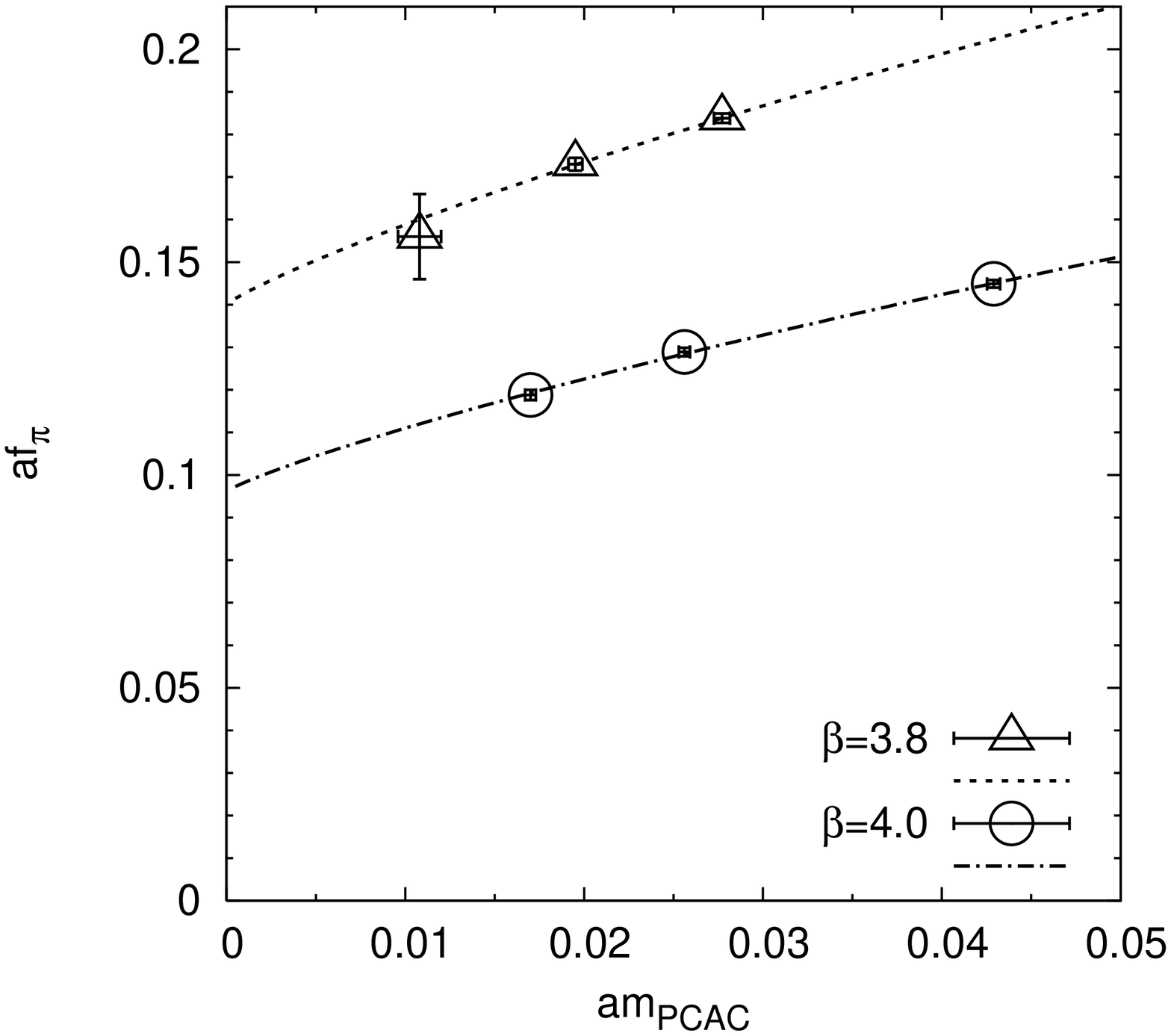}
  \caption{\label{fig:mpi}
    Pion masses squared and pion decay constants in lattice units and
    the results of the PQChPT fit.}
\end{center}
\end{figure}

The partially quenched ChPT formulae are used to extract the
corresponding low-energy coefficients from the pion data. 
Considering the number of lattice data at our disposal, a full fit
including all the terms in the ChPT formulae is not possible, so we
take only the continuum terms into account. We fitted the data for both $\beta$
values simultaneously
neglecting the dependence of the renormalizations factors
$Z_A$ and $Z_P$ upon the lattice coupling constant.  Introducing the one-flavor low-energy
constants
\begin{eqnarray}
\Lambda_3 &=& 4\pi F_0 \exp \{ 64 \pi^2 (L_4 + L_5 - 2L_6 - 2L_8) \}\,,
\nonumber \\
\Lambda_4 &=& 4\pi F_0 \exp \{ 64 \pi^2 (L_4 + L_5) \}\,,
\end{eqnarray}
the fit formulae of the renormalized values reduce to
\begin{eqnarray}
m_{\pi}^2 &=& \chi_{\rm\scriptscriptstyle PCAC}
+ \frac{\chi_{\rm\scriptscriptstyle PCAC}^2}{16 \pi^2 F_0^2}
\ln \frac{\chi_{\rm\scriptscriptstyle PCAC}}{\Lambda_3^2}\,,\nonumber\\
\frac{f^{\rm\scriptscriptstyle R}_{\pi}}{F_0 \sqrt{2}} &=&
1 - \frac{\chi_{\rm\scriptscriptstyle PCAC}}{32 \pi^2 F_0^2}
\ln \frac{\chi_{\rm\scriptscriptstyle PCAC}}{\Lambda_4^2}\,.
\end{eqnarray}
The data and the fitted curves are shown in
Fig.~\ref{fig:mpi}.

In oder to improve the numerical results for the universal low-energy
constants~$\Lambda_{3,4}$, which do not explicitly depend on the
lattice spacing~$a$, we also performed fits to the ratios~\cite{AlphaRatio,qq+q}
\begin{equation}
\frac{m_\pi^2}{m_{\pi,\rm ref}^2}\,,\quad
\frac{f_\pi}{f_{\pi,\rm ref}}\,.
\end{equation}
For this calculation we restricted ourself to the data at
$\beta=4.0$ with reference point at $\kappa=0.1615$. We obtain in this case the
results
\begin{eqnarray}
\frac{\Lambda_3}{F_0}&=&10.0 \pm 2.6\,,\\
\frac{\Lambda_4}{F_0}&=&31.5 \pm 14.3\,,
\end{eqnarray}
which, interestingly, are  compatible with phenomenological values obtained
from ordinary QCD~\cite{Duerr}. The errors are however  quite large 
(we hope to improve these determinations in the future).

In addition, we investigated the relation between the mass of the
pion and of the physical $\eta$, reducing to formula~(\ref{eq:eta}) at leading-order.
For this purpose we fitted simultaneously $m_\pi^2$ and $m_{\eta}^2$ as a function of 
the PCAC quark mass, again considering only $\beta=4.0$. This yields to
\begin{equation}
\alpha = -0.03(19)\,,\quad
am_\Phi = 0.18(8)\,,
\end{equation}
suggesting a vanishing $\alpha$. Using the value of $r_0/a$
extrapolated to vanishing PCAC quark mass and setting $\alpha=0$
we find
\begin{equation}
am_\Phi = 0.19(2)\quad{\rm or}\quad
r_0m_\Phi = 0.72(10)\,,
\end{equation}
which means
\begin{equation}
m_\Phi = 284\pm40\,\mathrm{MeV}
\end{equation}
in physical units.

The value of $m_\Phi$ can also be obtained from the
Witten-Veneziano formula~\cite{WittenVeneziano}
\begin{equation}
m_{\Phi}^2 = \frac{4 N_f}{(f_\pi^{\rm\scriptscriptstyle R})^2} \chi_t
\end{equation}
valid at leading-order in the ('t Hooft) large  $N_c$ limit. An estimate of the quenched
topological susceptibility present in the literature is
$\chi_t=(193\pm9\,\mathrm{MeV})^4$~\cite{chit}. Using our value for
$f_\pi^R$, which is subject to a sizeable statistical error, one would
obtain $m_\Phi = 450\pm170~\mathrm{MeV}$.

\section{Summary and outlook}

This first Monte Carlo investigation of $N_f=1$ QCD reveals the qualitative features
of the low lying hadron spectrum of this theory.
The lightest hadron is the pseudoscalar $\eta$ meson (see
Table~\ref{tab:hadmass} and Figure~\ref{fig:hadmass}) while
the scalar meson, the $\sigma$, is about a factor 1.5 heavier.
It is interesting to compare our data with the estimate in~\cite{1overN}
$m_{\sigma}/m_{\eta} \simeq N_c/(N_c-2)=3$ for $N_c=3$.
The above prediction applies for the massless theory and one could expect 
the agreement to improve for smaller quark masses.
Our bare quark masses (estimated from the PCAC quark mass in the valence analysis) range
between  10{\rm\,MeV} and 60{\rm\,MeV}, while the lightest pion mass is 
$\sim 270$~MeV.

The lightest baryon, the $\Delta$ $(\frac{3}{2}^+)$, 
is by about a factor 3 heavier than the $\eta$ meson.
The lightest scalar mass obtained with a glueball $0^{++}$ operator  
lies between the $\sigma$ meson and the
$\Delta$ baryon mass. However, this mass could be overestimated,
since, due the high level of noise, only small time-separations could be 
included in the analysis.

In general, the mass measurements have relatively large errors
between 3--10\%. In order to obtain more quantitative results,
larger statistics and  smaller quark masses are required.
We hope to be able to make progresses in both directions~\cite{OurProgr}
with our new simulations using Stout-smeared links in the fermion action.
Some preliminary results were already presented in this contribution
(see~\cite{StoutTM} for a test of this formulation in twisted mass QCD 
with $N_f=2$).

The introduction of a partially quenched extension of the single flavor theory
with valence quarks allows to define the bare quark mass in terms of the PCAC quark mass
of the fictitious multi-flavor theory. The computation of the bare quark mass is
intricate in the unitary theory due to the absence of a chiral symmetry (the arguments 
of~\cite{Creutz:UpMass} regard the definition of a {\em renormalized} quark mass). 
Comparison of lattice data with partially quenched  chiral perturbation theory
allowed the determination of some of the low-energy
constants of the chiral Lagrangian. The latter are compatible, even if with large error,
with recent lattice determinations for $N_f=2$ QCD.




A further direction of investigation for the future~\cite{OurProgr} is the CP-violating phase 
transition expected at negative quark masses~\cite{Creutz:CPNf1}. 
For this aspect of the single flavor theory
the non-positivity of the fermion measure plays an essential role.



\end{document}